\documentclass[showpacs,preprintnumbers,amsmath,amssymb,prl,twocolumn,
superscriptaddress,footinbib]{revtex4}
\usepackage{graphicx}% Include figure files
\usepackage{bm}% bold math
\usepackage{subfigure}
\def\comment#1{}

\begin{document}

\title{Metal-Insulator Transition in Two- and Three-Dimensional 
Logarithmic Plasmas}

\author{S. Kragset}
\email{steinar.kragset@phys.ntnu.no}
\affiliation{Department of Physics, Norwegian University of
Science and Technology, N-7491 Trondheim, Norway}
\author{A. Sudb{\o}}
\email{asle.sudbo@phys.ntnu.no}
\affiliation{Department of Physics, Norwegian University of
Science and Technology, N-7491 Trondheim, Norway}
\author{F. S. Nogueira}
\email{nogueira@physik.fu-berlin.de}
\affiliation{Institut f\"ur Theoretische Physik, Freie Universit{\"a}t Berlin,
Arnimallee 14, D-14195 Berlin, Germany}

\date{Received \today}

\begin{abstract}
  We consider the scaling of the mean square dipole moment in a plasma
  with logarithmic interactions in a two- and three-dimensional system. 
  In both cases, we establish the existence of a
  low-temperature regime where the mean square dipole moment does not
  scale with system size and a high-temperature regime where it does scale 
  with system size. Thus, there is a nonanalytic change in the
  polarizability of the system as a function of temperature, and hence
  a metal-insulator transition in both cases. The relevance of this
  transition in three dimensions to quantum phase transitions in
  $2+1$-dimensional systems is briefly discussed.
\end{abstract} 


\maketitle 

Gauge theories where matter fields are coupled to compact 
gauge-fields in $2+1$ dimensions have recently been much studied as 
effective theories for Mott insulators with competing 
orders \cite{Sachdev,Moessner,Senthil}. Compact $U(1)$ gauge theories in 
$2+1$ dimensions are capable of sustaining topological defects in the gauge 
fields in the form of space-time instantons \cite{Polyakov}. Such theories 
generically offer the possibility of featuring confinement-deconfinement 
transitions associated with a proliferation of topological defects in the gauge sector. 
It is hoped that phase transitions in matter coupled such theories 
\cite{FradkinShenker,EinhornSavit} may be connected to difficult problems in condensed 
matter physics, such as the breakdown of Landau Fermi liquid theory and possibly also 
spin-charge separation in strongly correlated systems at zero temperature in two spatial 
dimensions \cite{SenthilFisher}. One such model considered recently is the compact 
abelian Higgs model where the gauge field is coupled to matter fields with the 
{\it fundamental charge} \cite{KleiNogSu}. By itself, the compact $U(1)$ gauge 
sector may be mapped onto the $2+1$-dimensional Coulomb gas with $1/R$-interactions 
between ``point charges'', i.e., the instantonic topological defects of the theory 
\cite{Polyakov}. As shown by Polyakov, the latter system is always  in a  metallic phase 
(when the instantons are regarded as electric charges) \cite{Polyakov}. 
More recently, it was shown that when matter 
carrying the fundamental charge was coupled to the compact gauge sector, and critical 
matter field fluctuations were integrated out, the system could be mapped onto a
gas of point charges interacting with a potential $-\ln(R)$ in $2+1$ space-time
dimensions, instead of $1/R$ \cite{KleiNogSu}. Using RG arguments, it was demonstrated 
that the system with the $2+1$-dimensional logarithmic interaction may undergo a 
finite-temperature  phase transition driven by the unbinding of dipole 
configurations \cite{KleiNogSu}, from a low-temperature dielectric regime to a 
high-temperature metallic regime(again when the instantons are regarded as electric 
charges). Here, we demonstrate this using  Monte Carlo (MC) simulations.  

Whether or not there exists a low-temperature dielectric regime separated
from a high-temperature metallic regime in a three-dimensional system of point charges 
interacting with logarithmic interactions and overall charge-neutrality, is
presently a matter of debate in both the condensed matter and lattice gauge theory 
literature \cite{KleiNogSu,Herbut,Chernodub}. The theorem that dipoles cannot screen the 
Coulomb potential (in any dimension) \cite{Frohlich} does not apply to the case of a 
logarithmic pair potential between point charges in three dimensions. It is due to this 
inability of the dipoles to screen a Coulomb potential that the Kosterlitz-Thouless (KT) 
transition is possible: the low temperature dielectric phase is always critical below the 
KT transition temperature. In three dimensions, on the other hand, Debye-H\"uckel theory is 
essentially exact. Thus, starting from a system of charges, screening  is such that  the 
screening length cannot become infinite, 
and the system is always in the metallic phase. On the other hand, if one starts from a 
three-dimensional dipole system, the system always stays in the dielectric phase, since dipoles 
cannot screen. Once more, there is no phase transition. If some departure from the Coulomb 
potential occurs in three dimensions, dipoles could conceivably be able to screen. A $3D$ 
logarithmic interaction could in principle be screened to a potential that decays with 
separation between the charges \cite{Herbut}. If so, no metal-insulator transition 
occurs at any finite temperature, analogous to the situation in a three-dimensional 
Coulomb plasma  \cite{Polyakov}. If one thinks 
of the gas of point charges as a gas of instantonic defects in a compact gauge field, 
the lack of a metal-insulator transition due to non-standard screening by dipoles 
would correspond to permanent confinement in the compact gauge theory. 

We therefore consider a logarithmic plasma in two and three dimensions  on the lattice
using  MC simulations, and study the polarizability of the system as a function of 
temperature. The screening properties of these systems determine whether they are
insulators or metals and are governed by the dielectric constant $\varepsilon$, which 
in a low-density approximation is given by the polarizability $p$ of the system, $\varepsilon 
= 1 + n_d \Omega_d p$. Here $n_d$ is the dipole density, and $\Omega_d$ is the solid angle
in a $d$-dimensional system. Since the polarizability of the system is proportional to the 
mean square separation $\langle s^2 \rangle  \equiv \langle |\vec{r}_i - \vec{r}_j|^2 \rangle$ 
between the charges constituting the dipoles, it is natural to focus on 
$\langle s^2 \rangle$ in order to investigate whether the system is a 
dielectric or a metal. We find a low-temperature regime where dipoles are tightly bound, 
separated from a high-temperature regime where they are unbound. This implies 
the existence of a low-temperature insulating regime and a
high-temperature metallic regime, separated by a genuine 
phase transition. For comparison and benchmark purposes, we  compute the 
same quantities for the two-dimensional logarithmic plasma, where a 
metal-insulator transition in the form of a KT phase transition is known 
to exist \cite{KosterlitzThouless73,MinnhagenReview}.

%\begin{figure}[htbp]
%  \centering
%  \hspace{-0.5cm}
%  \scalebox{0.68}{\includegraphics{Potential.eps}}
%\caption{The potential $V(x,0,0)$ along the $x$ direction and with $L =
%  40$. The dashed line represents a fit of a periodic function $f(x) =
%  a[b + \ln x + \ln (40-x)]$ to $V$ via the parameters $a$ and
%  $b$, which resulted in $a = 1.161$ and $b = -2.385$.}
%  \label{fig:potential}
%\end{figure}
We consider the Hamiltonian
\begin{equation}
  \label{eq:Hamiltonian}
  \mathcal{H} = \frac {1}{2} \sum_{i,j} q_i V(\vec{r}_i -
  \vec{r}_j) q_j , %Sjekk prefaktorer!!!
\end{equation}
with 
\begin{equation}
  \label{eq:potential}
  %\begin{split}
    V(\vec{r})  = \frac{4 \pi^2}{Na^d}\sum_{\vec{k}} 
    \frac {[e^{i\vec{k}\cdot \vec{r}} - 1]}{[2(d - 
\sum_{\alpha=1}^d\cos k_\alpha a)]^{d/2}},
  %\end{split}
\end{equation}
where the sum is over all pairs of sites of a two- or three-dimensional periodic lattice 
and $q_i$ is the charge at site $i$, $N = L^d$ is the total number of sites, and $a$ is 
the lattice spacing. We have subtracted $V(\vec{r}=0)$ from the potential, since
only neutral configurations will contribute to the partition function, and 
we work at zero chemical potential for the instantons \cite{LeeTeitel}. This has 
the usual Coulomb form for $d=2$, but differs from it in the $d=3$ case:  instead of the 
usual $1/R$ potential of the three-dimensional Coulomb gas ($3D$CG), the power $3/2$ 
results in a logarithmic potential.
%as can clearly be seen in Fig. \ref{fig:potential}, very
%much alike the potential of the two-dimensional Coulomb gas ($2D$CG). 
%Furthermore, we can 
Taking the continuum limit of Eq. (\ref{eq:potential}) for $d=3$,
we find
%in the form
\begin{eqnarray}
V(\vec{r})=\int\frac{d^3k}{(2\pi)^3}\frac{[e^{i\vec{k}\cdot\vec{r}}-1]}{
(k^2+m^2)^{3/2}}~~~~~~~~~~~~~~\nonumber\\
=\frac{1}{2\pi^2}\left[K_0(mr)+\frac{\Lambda}{\sqrt{\Lambda^2+m^2}}
-{\rm arcsinh}\left(\frac{\Lambda}{m}\right)\right],
\end{eqnarray} 
where $K_0(x)$ is a Bessel function, $\Lambda=2\pi/a$, and the Debye-H\"uckel 
screening length $m^{-1}$ was introduced as an infrared regulator. As $m\to 0$ 
we have
\begin{equation}
V(r)=\frac{1}{2\pi^2}[1-\gamma-\ln(\Lambda r)]+{\cal O}\left(\frac{m}{\Lambda}\right)^2,
\end{equation}
where $\gamma$ is the Euler constant. Therefore, Eq. (\ref{eq:potential}) 
behaves exactly in the same way as the two-dimensional potential. 

Let us first turn to the $2D$CG which we use as a benchmark on our 
method of probing the metal-insulator transition, before proceeding
to the three-dimensional logarithmic gas ($3D$LG). 
%which has the
%Hamiltonian given in Eq. (\ref{eq:Hamiltonian}), but now with the
%sum over a two-dimensional $L \times L$ lattice 
%\begin{equation}
%  \label{eq:2DCGpotential}
%    V(\vec{r}_i - \vec{r}_j)  =  \frac{4 \pi^2}{N} \sum_{\vec{k}} 
%    \frac {e^{\vec{k}\cdot\vec{r}} - 1}{2[2 - \cos(k_x) - \cos(k_y)]}.
%\end{equation}
At low temperatures this model consists of tightly bound dipoles with
the two opposite charges separated by a distance $|\vec{r}_i -
\vec{r}_j| \equiv s$ of the order the lattice constant. 

In a $2D$ system, it is well-known that dipoles begin to unbind at a
critical temperature $T_{KT}$ \cite{KosterlitzThouless73} and at high 
temperatures the $2D$CG is a fully ionized metallic plasma,
separated from the low-temperature dielectric insulating phase by a 
genuine KT phase-transition. Thus, we expect no finite-size scaling of 
$\langle s^2 \rangle$ below the KT-transition, whereas we should expect 
$\langle s^2 \rangle \propto L^{\alpha(T)}$ with $\alpha(T) \leq 2$ at 
higher temperatures. Using an intuitive low density argument, neglecting 
screening effects \cite{KosterlitzThouless73}, we can calculate the 
behaviour of $\langle s^2 \rangle$ to leading order in $L$,
\begin{equation}
  \label{eq:ssquared}
  \langle s^2 \rangle \propto 
    \begin{cases} {\rm Const.} & ;  \; \;   T < T_{KT} \\
                          L^{ (T - T_{KT})/T } & ;  \; \; T_{KT} < T < 2 T_{KT} \\
                          L^2 & ;  \; \;  2 T_{KT} < T.
    \end{cases}
\end{equation}
Hence, $\alpha(T)$ is zero for low temperatures and a monotonically
increasing function of temperature just above $T_{KT}$. 

Including screening effects in $2D$ shows that this conclusion still holds, 
however the temperature at which it occurs is determined by screening. This 
is ultimately related to the fact that dipoles contribute only through a 
correction to the coefficient of the bare logarithmic pair-potential. Hence, 
in $2D$ the functional form of the renormalized potential is unaltered, only 
prefactors are changed. In three dimensions, it is far from obvious that dipoles 
do not have a much more disruptive effect on a bare logarithmic pair-potential.
 
To adress the issue of scaling of $\langle s^2 \rangle$, we will use MC 
simulations and finite-size scaling to study $\langle s^2 \rangle$ 
both in the $2D$CG as well as in the  $3D$LG. 
In our simulations, the 
particle number is not conserved. However, during the simulations the system 
is maintained electrically neutral. The MC moves involve creation 
and annihilation of charges, applying the Metropolis algorithm in this
process. Starting at some randomly chosen lattice site, an attempt 
to insert a negative or positive charge at random at this site is made,
with an opposite charge at a nearest-neighbour site. The move is accepted 
with probability $\exp (-\Delta E / T) = \exp [-(\mathcal{H_{\rm{new}}} -
\mathcal{H_{\rm{old}}})/T]$, and this is done in all $d$ directions,
before we move to the next site. It is clear that placing a charge on top 
of an opposite one corresponds to the annihilation of the existing one. In 
order to measure $s$ we have to keep track of which two charges belong to
each other in a dipole, since there is no physical link defining the
dipole. Two charges inserted into the lattice at the same MC
move is chosen as a dipole, and if one of these charges subsequently is
annihilated, the effect of this annihilation is to diffuse one of the
charges in the dipole. Hence, such a diffusion may increase or decrease 
$s$ for this dipole \cite{footnote}. One sweep is defined as going through 
all the sites in the lattice once, and at every  tenth sweep we sample $s^2$
averaged over the system. The sample is, however, rejected if there are no 
charges in the system at this MC time, since there is no information 
on $\langle s^2 \rangle$ in such a configuration. The thermal average obtained 
at the end of the simulation is thus taken only over non-empty configurations. 

\begin{figure}[htbp]
  \centering
  \scalebox{0.7}{
    \hspace{-.8cm}
    \includegraphics{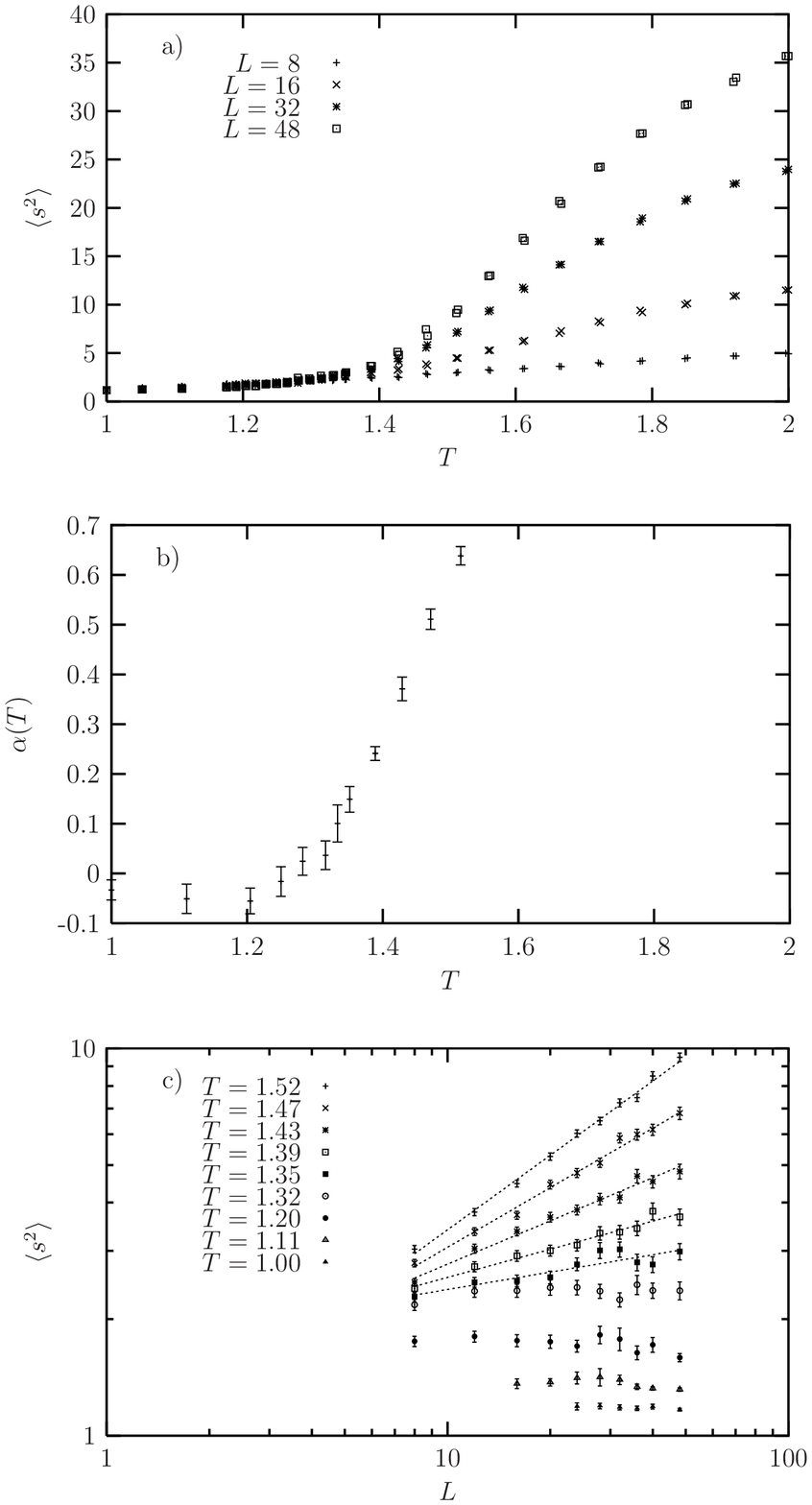}
  }
  \caption{Results from MC simulations of the $2D$CG, where 
    $1.0 \cdot 10^5$ sweeps were used at each temperature. a) 
    $\langle s^2 \rangle$ 
    versus $T$ for a selection of the 
    simulated system sizes $L =$8, 12, 16, 20, 24, 28, 32, 36, 40 and
    48. Errorbars are smaller than the symbols used. b) $\alpha$ versus $T$
    found from fitting the data of $\langle s^2 \rangle$ at different
    $T$ to $A L^{\alpha}$ where $A$ is a constant. A selection of such fits 
    are shown in c). We note that $\langle s^2 \rangle$ is practically
    independent of $L$ up to a certain $T$.}
  \label{fig:2dcgresults}
\end{figure}

The results from the simulations on the $2D$CG are presented in Fig.
\ref{fig:2dcgresults}. A total of $10^5$ MC sweeps at each
temperature are used to produce the first plot. From the $\langle s^2
\rangle$-data of system sizes $L =$ 8, 12, 16, 20, 24, 28, 32, 36, 40
and 48, we extract $\alpha$ for temperatures in the interval $(1.0,
1.52)$, which is shown in the second plot. It is evident that
there are two distinct regimes of temperatures, one in which the
charges of almost all dipoles are bound as tightly as possible, the
separation of the charges correspond to the lattice constant. In the
high-temperature regime the dipoles have started to separate,
reflected by a scaling of $\langle s^2 \rangle \sim L^{\alpha(T)}$ with 
the system size. The two regimes are necessarily separated by a phase transition,
since in the low-temperature regime $\alpha(T)=0$ while in the
high-temperature regime $\alpha(T) \neq 0$. This necessarily implies
a non-analytic behavior of $\alpha(T)$. An attempt to determine the transition 
temperature from these plots yields approximately $1.32$. This is slightly less 
than  the early results of Saito and M\"{u}ller-Krumbhaar of $1.35$ in our units
\cite{SMK}, but is in excellent agreement with much more recent simulations by 
Olsson \cite{Olsson}. It provides confidence in the method of locating the 
critical point by monitoring the quantity $\langle s^2 \rangle$, even
when the system is subjected to periodic boundary conditions \cite{footnote}. 

\begin{figure}[htbp]
    \scalebox{0.7}{
      \hspace{-0.8cm}
      \includegraphics{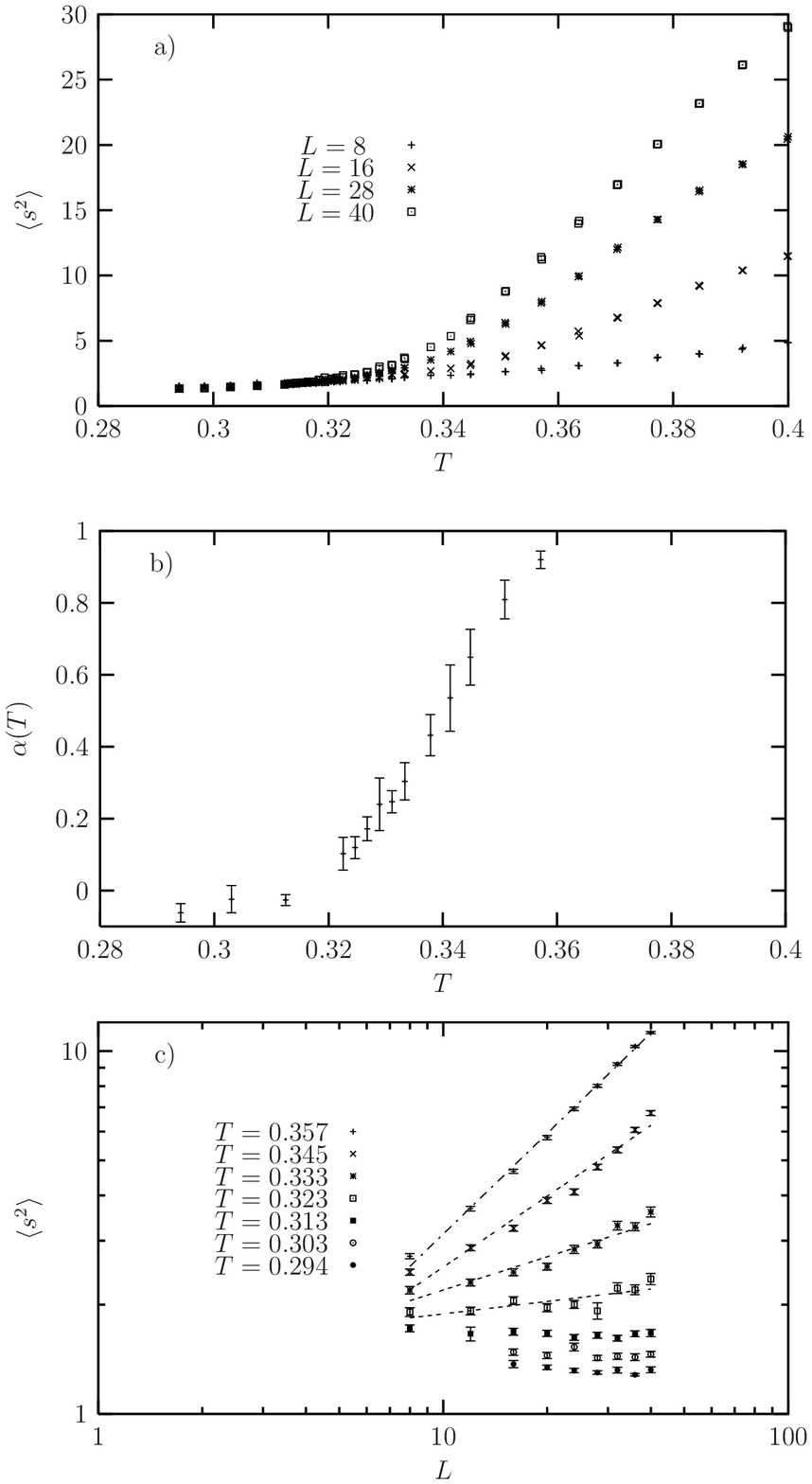}
    }
\caption{Results from simulations of the $3D$LG. Up to
  $2.0 \cdot 10^5$ MC sweeps were used at each temperature. 
  a) $\langle s^2 \rangle$ vs $T$ for some of the 
  simulated system sizes $L =$8, 12, 16, 20, 24, 28, 32, 36 and
  40. Error bars are smaller than the symbols used. b) $\alpha(T)$ vs $T$
  found from fitting the data of $\langle s^2 \rangle$ at different
  $T$ to $A L^{\alpha}$ where $A$ is a constant. A selection of such 
  fits are shown in c). It is evident that a low-temperature regime 
  where $\langle s^2 \rangle$ is independent of $L$ exists in the 
  $3D$ log-gas in the same that that it exists in the $2D$ log-gas.}
 \label{fig:3dloggasresults}
\end{figure}

Exactly the same simulation technique is applied to the $3D$LG
and the results are shown in Fig. \ref{fig:3dloggasresults}. The
finite-size analysis of $\langle s^2 \rangle$ in order to extract
$\alpha$ is here done on the basis of system sizes $L =$8, 12, 16, 20,
24, 28, 32, 36 and 40 and up to $2.0 \cdot 10^5$ MC sweeps are
used. As in the $2D$CG we see that the system exhibits two distinct
regimes, one insulating regime consisting of tightly bound dipoles and
one metallic. There is no scaling of $\langle s^2 \rangle$ with $L$ below 
$T \simeq 0.32$ with the system size. Note that the change in scaling of 
the mean square dipole moment occurs at a significantly lower temperature 
than in the $2D$CG. This is to be expected, since there is more configurational 
entropy available in $3D$ than in $2D$. It is worthwhile comparing this 
result with the one obtained in Ref. \cite{KleiNogSu} where the 
coupling $4\pi^2/t$ corresponds to the temperature $T$ here. There 
 the critical value $t_c=12\pi^2$ was obtained, corresponding to 
$T_c=1/3$ in our case and agreeing well with our numerical result.  

The main result is that in 
the $3D$LG and the $2D$CG, a low-temperature regime exists where positive 
and negative charges are bound in tight dipole pairs. This regime is separated 
from a high-temperature regime where at least a finite fraction of charges are 
free. The results obtained in three dimensions have the same features as those 
found in $2D$. The scaling exponent $\alpha(T)$ for $\langle s^2 \rangle$ is 
zero in the low-temperature phase and positive in the high-temperature 
phase. Such a change in $\alpha(T)$ cannot be analytical, and therefore the 
two scaling regimes must be separated from each other via a phase transition. 
Since  $\langle s^2 \rangle$ is a measure of the polarizability of the system, 
we conclude that the above demonstrates a non-analytic change in the polarizability 
of the system, i.e., a non-analytic change in the dielectric function of the system 
as a function of temperature. Hence, a system of point charges with overall charge 
neutrality interacting with a bare logarithmic interaction undergoes a metal-insulator 
transition in both $2D$ and $3D$. 

The $3D$LG can be shown to be equivalent to an anomalous sine-Gordon theory in $d=3$   
where the usual $k^2$ dispersion is replaced by a $|k|^3$ one \cite{KleiNogSu}, which 
is non-analytic. This leads to  technical difficulties in a standard renormalization 
group (RG) calculation, where in a perturbative treatment only {\it analytic} 
singularities are generated. Therefore, in a standard RG analysis an analytic 
correction $\sim k^2$ to the dispersion can be generated. On the other 
hand, since  $|k|^3$ is non-analytic, it cannot be renormalized within a 
standard RG analysis. It is conceivable that this is one of the reasons why a 
recent RG analysis of this theory did  not find evidence for a dielectric phase 
\cite{Herbut}. 

We emphasize that although it is known that a metal-insulator transition occurs in 
$2D$ via a KT transition \cite{KosterlitzThouless73}, our numerics 
by themselves do not demonstrate this. To establish the KT nature of 
the transition  on purely numerical grounds requires convincing numerical evidence 
that there is a universal jump in the  inverse dielectric constant at the transition. 
The main point of the present simulations is that they settle the difficult question 
of whether a low-temperature dielectric regime exists at all in a $3D$LG, which is a 
system  not subject to standard electrostatics, and for which the usual theorems on 
screening of charges by dipoles do not apply. The answer is in the affirmative, and we 
have applied our method also to the $2D$ case for comparison and as a benchmark on 
the correctness of method of monitoring $\langle s^2 \rangle$ to establish the existence 
of two scaling regimes separated by a phase transition.       
 
The authors acknowledge stimulating discussions and correspondence with I. F. Herbut 
and S. Sachdev. This work received financial support from the Norwegian University of 
Science and Technology through the Norwegian High Performance Computing Program (NOTUR), 
from the Research Council of Norway, Grants No. 158518/431, 158547/431 (NANOMAT), 
157798/432 (A.S), and from DFG Priority Program SPP 1116 (F.S.N.). One of us (S.K.) 
thanks the Norwegian University of Science and Technology for a PhD Fellowship.

%\bibliography{master}

\begin{thebibliography}{100}
\bibitem{footnote} Periodic boundary conditions are enforced by decomposing the
dipole-vectors in cartesian components $s_\alpha, \alpha \in (x,y,z)$ 
and letting $s_\alpha \to L - s_\alpha$ if $s_\alpha > L/2$.
\end{thebibliography}

\end{document}